\documentclass[fleqn,twoside]{article}
\usepackage{espcrc2}

\usepackage{epsfig}

\usepackage{graphicx}
\usepackage[figuresright]{rotating}

\newcommand{\AmS}{{\protect\the\textfont2
  A\kern-.1667em\lower.5ex\hbox{M}\kern-.125emS}}

\title{How light can the Higgs be?}

\author{K.~Holland\address[UCSD]{Department of Physics,
        University of California at San Diego, La Jolla CA 92093-0319, USA}%
        \thanks{Speaker at the conference. Research supported by the DOE
        under grant DOE-FG03-97ER40546.}
        and J.~Kuti\addressmark[UCSD]}
       
\begin{document}

\begin{abstract}
It is widely believed that, for a given Top mass, the Higgs mass has a lower
bound: if $m_{\rm Higgs}$ is too small, the Higgs vacuum is unstable due to 
Top dynamics. From vacuum instability, the state-of-the-art calculation of 
the lower bound is close to the current experimental limit. Using 
non-perturbative simulations and large $N$ calculations, we show that the 
vacuum is in fact never unstable. Instead, we investigate the existence of a 
new lower bound, based on the intrinsic cut-off of this trivial theory.
\end{abstract}

\maketitle

\section{INTRODUCTION}

It is generally accepted that the Higgs sector of the Standard Model
is trivial. A quantum field theory is defined by a set of bare parameters
and a regulator with some cut-off $\Lambda$. Renormalized quantities are
calculated in terms of the bare parameters, then the cut-off is sent to
infinity. In a trivial theory, the renormalized couplings vanish as
$\Lambda \rightarrow \infty$ for any choice of bare parameters. To have a 
non-trivial interacting theory, the cut-off must remain finite. Hence, the 
Higgs sector is an effective theory which is only valid at energy scales 
below the cut-off $\Lambda$. 

\begin{figure}[htb]
\epsfig{file=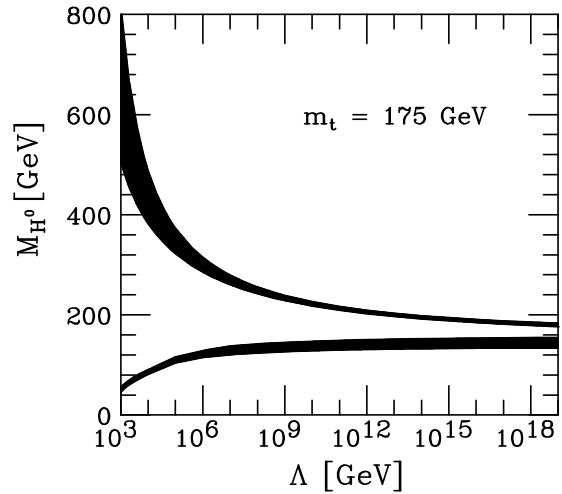,width=6.5cm,angle=90}
\vspace*{-0.75cm}
\caption{The phenomenological upper and lower bounds for 
	$m_{\rm Higgs}$, from \cite{Hagiwara:fs}.}
\label{fig:PDG_higgs}
\vspace*{-0.9cm}
\end{figure}

Fig.~\ref{fig:PDG_higgs} shows the current phenomenological upper and lower 
bounds for $m_{\rm Higgs}$ as a function of $\Lambda$, the threshold of new
physics \cite{Hagiwara:fs}. For a given cut-off, the upper bound is the 
largest $m_{\rm Higgs}$ that can be generated by any choice of bare 
parameters. Lattice simulations without Top dynamics have found that 
$m_{\rm Higgs} \simeq 650~{\rm GeV}$ when the bare Higgs coupling 
$\lambda \rightarrow \infty$ and the momentum cut-off $\Lambda = \pi/a$ is 
a few TeV \cite{Kuti:1987nr,Luscher:1988uq}, where $a$ is the lattice spacing.
For the cut-off effects to be acceptably small, we require the correlation 
length $\xi/a = 1/(m_{\rm Higgs}a) \geq 2$ so that e.g.~violation of 
rotational symmetry in particle scattering is less than a few \% 
\cite{Luscher:1988uq}.
%

The phenomenological lower bound for $m_{\rm Higgs}$ is based on the 
belief that, for a given Top quark mass, the Higgs mass cannot be too small, 
otherwise the Higgs field effective potential $V_{\rm eff}(\phi)$ is 
unstable. The one-loop contributions to the renormalized effective potential 
from Higgs and Top loops are of the form $\lambda_{\rm R}^2 \phi_{\rm R}^4 
\ln(\phi_{\rm R}/v_{\rm R})$ and $-y_{\rm R}^4 \phi_{\rm R}^4 
\ln(\phi_{\rm R}/v_{\rm R})$ respectively, where $v_{\rm R}$ 
is the renormalized vacuum expectation value of $\phi_{\rm R}$, and 
$\lambda_{\rm R}$ and $y_{\rm R}$ are the renormalized Higgs and Yukawa
couplings. If $y_{\rm R}^2 \gg \lambda_{\rm R}$, the negative Top contribution
dominates, $V_{\rm eff} \rightarrow -\infty$ as $\phi_{\rm R} 
\rightarrow \infty$ and the potential no longer has its absolute minimum at
$v_{\rm R}$. For a given $m_{\rm Higgs}$ and $m_{\rm Top}$ 
(i.e.~$\lambda_{\rm R}$ and $y_{\rm R}$), renormalized perturbation theory
is valid only if the vacuum is stable, defining an energy scale 
$\Lambda$ where new physics must occur. 

The state-of-the-art calculation of the lower bound from vacuum instability
gives $m_{\rm Higgs} \geq 86~{\rm GeV}$ for $\Lambda = 10~{\rm TeV}$ and
$m_{\rm Top} = 175~{\rm GeV}$, with an uncertainty of less than $5~{\rm GeV}$
\cite{Casas:1996aq}. As the current experimental limit is 
$m_{\rm Higgs} \geq 114.3~{\rm GeV}$, this is a very relevant and 
important statement for the validity of the Standard Model. However, we claim 
that there is no vacuum instability in the theory, making the validity
of the current lower bound uncertain.

\section{THERE IS NO INSTABILITY}

It was first shown in \cite{Kuti:bs} how to measure the constraint effective 
potential $V_{\rm eff}$ non-perturbatively via lattice simulations. In 
particular, we examine a model of a single component real scalar field 
coupled to two copies of staggered fermions, corresponding to 8 degenerate
continuum fermion flavors. The bare parameters are the Higgs mass $m$ and the 
Higgs and Yukawa couplings $\lambda$ and $y$. The scalar field is coupled 
locally to the fermions. We use the leapfrog Hybrid Monte Carlo algorithm 
with step-size $\Delta t = 0.01$ and trajectory length $N_t \Delta t \geq 1$. 
Our statistics are $10^4$ trajectories, with a separate simulation required 
at every value of the scalar field to measure $V_{\rm eff}$. Further details 
will be presented in 
\cite{Holland03}.

\begin{figure}[htb]
\vspace*{-1.25cm}
\epsfig{file=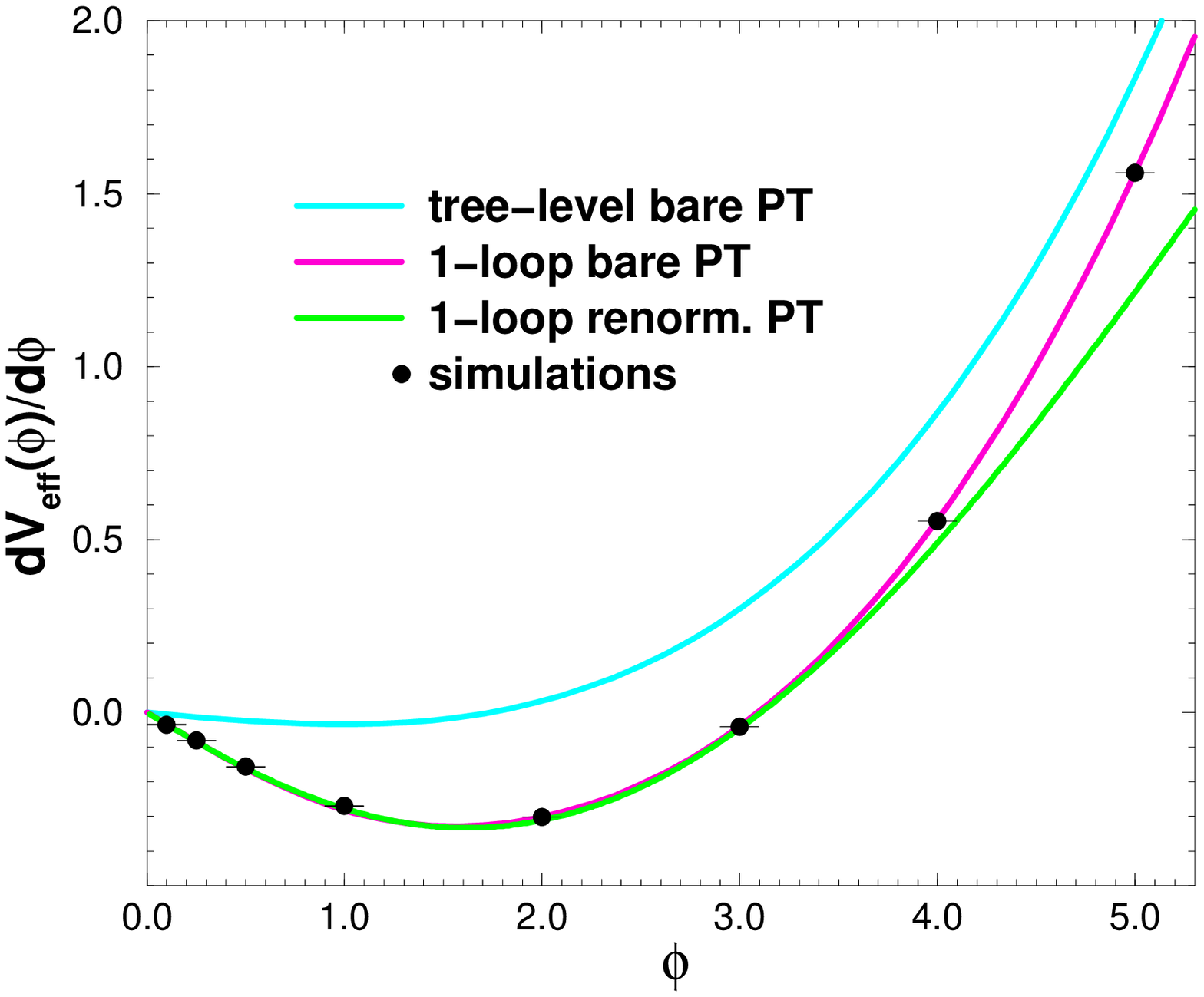,width=8cm,angle=0}
\vspace*{-1.75cm}
\caption{The derivative of the constraint effective potential 
         $dV_{\rm eff}(\phi)/d\phi$.}
\label{fig:potential}
\vspace*{-0.75cm}
\end{figure}

Fig.~\ref{fig:potential} shows results for the derivative of the 
constraint effective potential $dV_{\rm eff}(\phi)/d\phi$ for a particular
choice of bare parameters with $y^2 \gg \lambda$. There is no sign of any 
instability at large values of the scalar field. One-loop bare perturbation 
theory is in excellent agreement with the non-perturbative simulations, and 
when $\phi$ is large, even tree-level perturbation theory is very close to the 
full result. 

\begin{figure}[htb]
\vspace*{-0.5cm}
\epsfig{file=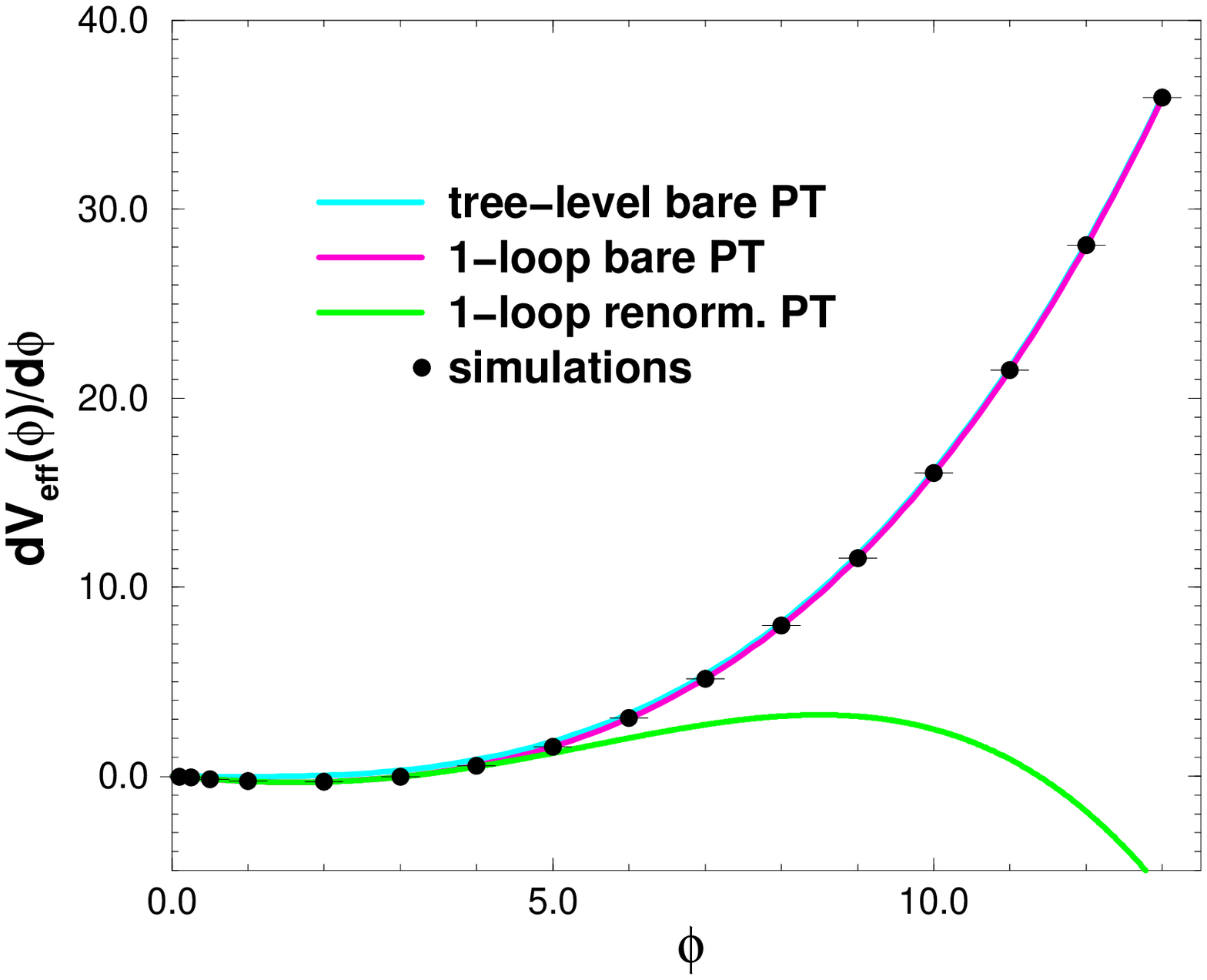,width=8cm,angle=0}
\vspace*{-1cm}
\end{figure}

Renormalized perturbation theory also agrees with the non-perturbative
determination of $V_{\rm eff}$ when $\phi$ is small. However, when $\phi$ is
large, the renormalization method breaks down because the cut-off terms
${\cal{O}}(\phi^2/\Lambda^2)$ are not small. The effective potential
appears unstable in renormalized perturbation theory because these large 
positive cut-off effects are neglected. The effective potential is always 
stable, as shown by numerical simulations and cut-off-dependent bare 
perturbation theory. This picture is confirmed in a calculation of the 
effective potential using a Pauli-Villars regulator, in the limit of a large 
number of fermion flavors \cite{Holland03}. 

\section{CORRECT LOWER BOUND}

In the same model of a single component real scalar field coupled to 
staggered fermions, we show how to compute the correct $m_{\rm Higgs}$ 
lower bound (see \cite{Lin:1993hp} for some previous studies of
Higgs--Top systems). We choose the bare parameters $\lambda, y$ and $m$ to be 
in the Higgs phase of the theory, close to the critical surface which 
separates the Higgs and the symmetric phases. For a particular choice of bare 
parameters, we measure $v_{\rm R}a$, the renormalized vev in lattice
spacing units, using $v_{\rm R}=246~{\rm GeV}$ to convert the momentum
cut-off $\Lambda=\pi/a$ into physical units. 

\begin{figure}[htb]
\vspace*{-0.6cm}
\epsfig{file=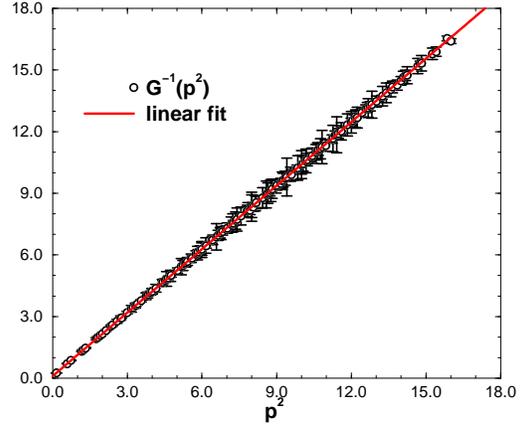,width=8cm,angle=0}
\vspace*{-1.75cm}
\caption{The inverse propagator $G^{-1}_{\phi}(p^2)$.}
\label{fig:propagator}
\vspace*{-0.8cm}
\end{figure}

We measure $m_{\rm Higgs}$ and the wave-function renormalization factor
$Z_\phi$ from the scalar field propagator in momentum space $G_\phi(p^2)$
\begin{equation}
G^{-1}_\phi(p^2) = (m^2_{\rm Higgs} + \hat{p}^2)/Z_\phi,
\end{equation}
\label{eq:propagator}
where $\hat{p}^2 = \sum_\mu 4\sin^2(p_\mu/2)$. We measure $m_{\rm Top}$ from
the zero momentum piece of the fermion propagator. All simulations were 
performed in $8^3 \times 16$ volumes, again using the HMC algorithm, varying 
the step-size $\Delta t$ to achieve acceptance rates of more than $90\%$. Our
statistics are $10^4$ trajectories for each choice of bare parameters.
In Fig.~\ref{fig:propagator}, we show a typical measurement of 
$ G^{-1}_\phi(p^2)$ giving $m_{\rm Higgs}a=0.310(3)$ and $Z_\phi=0.971(2)$.

\begin{figure}[htb]
\vspace*{-0.5cm}
\epsfig{file=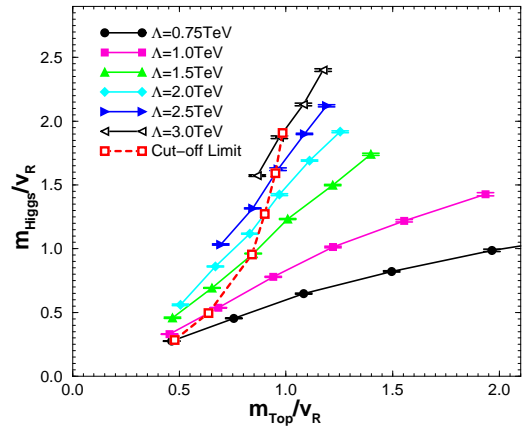,width=8cm,angle=0}
\vspace*{-1.75cm}
\caption{$m_{\rm Higgs}/v_{\rm R}$ vs.~$m_{\rm Top}/v_{\rm R}$
	 for various $\Lambda$.}
\label{fig:lower_bound}
\vspace*{-0.75cm}
\end{figure}

In Fig.~\ref{fig:lower_bound}, we plot a summary of the simulations.
The solid lines are measurements with $\Lambda$ fixed in physical units,
varying the bare parameters to stay at a fixed distance from the critical
surface. For a given $\Lambda$ and $m_{\rm Top}$, the smallest Higgs mass
is generated when $\lambda \rightarrow 0$. This is completely analogous to 
the upper bound, where $m_{\rm Higgs}$ is largest when $\lambda  
\rightarrow \infty$. To keep the cut-off effects acceptably small, we 
require both the scalar and fermion correlations lengths $\xi/a \geq 2$. 
The dashed line in Fig.~\ref{fig:lower_bound} corresponds to $\xi/a=2$, to the
left of the dashed line is the allowed region of small cut-off effects. From
this, we can extract the lower bound. For example, in this model of 8 
degenerate fermions, the smallest Higgs mass that can be generated for 
$m_{\rm Top}=175~{\rm GeV}$ and $\Lambda=2~{\rm TeV}$ is 
$m_{\rm Higgs} \simeq 230~{\rm GeV}$. 

\section{SUMMARY}

The apparent instability of the Higgs vacuum when coupled to a heavy Top 
quark is due to the breakdown of renormalized perturbation theory when the 
cut-off effects are large. This places the current phenomenological 
$m_{\rm Higgs}$ lower bound in doubt. The correct lower bound in a particular 
cut-off scheme can be found by measuring the smallest Higgs mass that can be 
generated while simultaneously keeping the cut-off effects acceptably small. 
A similar calculation for a realistic Higgs--Top system is a very timely and 
important project for the lattice community.


\begin{thebibliography}{9}

\bibitem{Hagiwara:fs}
K.~Hagiwara {\it et al.}  [Particle Data Group Collaboration],
Phys.\ Rev.\ D {\bf 66}, 010001 (2002).

\bibitem{Kuti:1987nr}
J.~Kuti, L.~Lin and Y.~Shen,
Phys.\ Rev.\ Lett.\  {\bf 61}, 678 (1988);
G.~Bhanot and K.~Bitar,
Phys.\ Rev.\ Lett.\  {\bf 61}, 798 (1988);
A.~Hasenfratz and T.~Neuhaus,
Nucl.\ Phys.\ B {\bf 297}, 205 (1988);
U.~M.~Heller {\it et al.},
Nucl.\ Phys.\ B {\bf 405}, 555 (1993).

\bibitem{Luscher:1988uq}
M.~L\"{u}scher and P.~Weisz,
Nucl.\ Phys.\ B {\bf 318}, 705 (1989).

\bibitem{Casas:1996aq}
J.~A.~Casas, J.~R.~Espinosa and M.~Quiros,
Phys.\ Lett.\ B {\bf 382}, 374 (1996).

\bibitem{Kuti:bs}
J.~Kuti and Y.~Shen,
Phys.\ Rev.\ Lett.\  {\bf 60}, 85 (1988).

\bibitem{Holland03}
K.~Holland and J.~Kuti, in preparation.

\bibitem{Lin:1993hp}
L.~Lin {\it et al.},
Phys.\ Lett.\ B {\bf 317}, 143 (1993);
Y.~Shen {\it et al.},
Nucl.\ Phys.\ B (Proc.\ Suppl.) {\bf 9}, 999 (1989);
W.~Bock {\it et al.},
Nucl.\ Phys.\ B {\bf 400}, 309 (1993).


\end{thebibliography}
\end{document}